\documentclass[cits]{PoS}

\title{Neutral mesons and disconnected diagrams in Twisted Mass QCD}

\ShortTitle{Neutral mesons and disconnected diagrams in Twisted Mass QCD}

\author{\speaker{Chris Michael}%
         \thanks{On behalf of the ETM Collaboration}\\
        Theoretical Physics Division, Dept of Mathematical Sciences,
 University of Liverpool, Liverpool L69 3BX, UK\\
        E-mail: \email{c.michael@liv.ac.uk}}

\author{Carsten Urbach\\
        Theoretical Physics Division, Dept of Mathematical Sciences, 
 University of Liverpool, Liverpool L69 3BX, UK\\
        E-mail: \email{c.urbach@liv.ac.uk}}

\abstract{

   \begin{center}
      \includegraphics[draft=false]{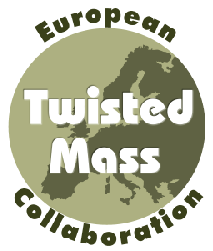}
    \end{center}
  \vskip 0.5cm

 We evaluate properties of neutral mesons in $N_f=2$ dynamical
simulations of TMQCD at maximal twist. The pion is explored - 
establishing the size of the isospin splitting (an order $a^2$ effect).
We investigate the $\eta'$ (the $N_f=2$ flavour singlet pseudoscalar
meson) and neutral  $\rho$ and scalar mesons. We  show that disconnected
diagrams can be evaluated very efficiently in TMQCD  using variance
reduction methods. 
}

\FullConference{The XXV International Symposium on Lattice Field Theory\\
		 July 30 - August 4 2007\\
		 Regensburg, Germany}

\usepackage{graphicx}

\begin{document}

\section{Introduction}

 Here we discuss TMQCD at maximal twist with $N_f=2$ degenerate sea
quarks using configurations from
ETMC~\cite{Boucaud:2007uk,Urbach:2007aa,ETMClong}. In particular we 
focus on neutral mesons, which have some unusual properties in TMQCD. 
Thus we evaluate the  disconnected contributions which are needed for a
study of neutral mesons~\cite{Jansen:2005cg}. We present a new method 
which, for twisted mass, allows  many disconnected contributions to be
evaluated very efficiently. 

 We then present results for pions (exploring the charge splitting
between $\pi^0$ and $\pi^+$). We also present  results for flavour
singlet pseudoscalar mesons (the $\eta_2$ meson), for vector mesons
(charge splitting and  decays), and for flavour-singlet scalar mesons.

\section{Disconnected diagrams - variance reduction}

 TMQCD has a degenerate pair of $u,\ d$ quarks with fermion matrix:
\ \ \ $ M_{u,d}= M_W \pm  i \mu \gamma_5  $ where $M_W$ is the Wilson-Dirac 
matrix. 
 Hence $  1/M_u - 1/M_d =  -2i\mu (1/M_d) \gamma_5  (1/M_u) $.

 Consider the disconnected contribution  $ \sum X (1/M_u - 1/M_d)$ where
 $X$ is some $\gamma$-matrix and/or colour-matrix and  the sum  is over
space. The conventional method involves solving
 $  \phi_r=(1/M_u) \xi_r$ with  stochastic volume sources $\xi_r$, then 
 $ \sum X/M_u  = \sum \langle \xi^* X \phi \rangle_r$ where the average
is over noise samples (labelled $r$).
 However, the case mentioned above can be  evaluated efficiently using
the `one-end-trick'~\cite{Foster:1998vw,McNeile:2006bz}. Then the
required disconnected loop is given by
 $$
  \sum X (1/M_u - 1/M_d) = 
    -2i\mu \sum \langle  \phi^* X \gamma_5 \phi \rangle_r
 $$

This has 
 signal/noise  $V/\sqrt{V^2}=1 $ which is much more  favourable than the
conventional method with signal/noise $1/\sqrt{V}$ (here $V=L^3 T$).

  For example,  at $\beta=3.9$, $\mu=0.004$, $L=24$, 
 (where  $M(\pi)\approx 300$ MeV, $a=0.086$ fm, $La=2.1$ fm) taking
$X=i$ which is appropriate for the $\eta_2$ correlator, we evaluate the
momentum-zero loop at a time-slice. We find  a standard deviation of 
$\sigma=18$ from inherent gauge-time variation whereas the  stochastic
noise is $\sigma=87$ from 24 samples of volume source (conventional
method) but  only  $\sigma=7.5$ from above method (with 12 samples).
 So 12 inversions give the disconnected correlator from all $t$ to  all
$t'$ with no significant increase in  errors from the stochastic
evaluation.

  For cases where this method cannot be used, 
eg. $\pi^0$ with $\bar{\psi}
\gamma_5 \tau_3 \psi \to \bar{\chi} I \chi$ at maximal twist,
 we use hopping  parameter variance reduction~\cite{McNeile:2000xx}
instead.

\section{Pion order($a^2$) effects}

\begin{figure}[thb]
\begin{center}
   \begin{tabular}{l@{\hspace{1cm}}l}
        \includegraphics[width=7cm]{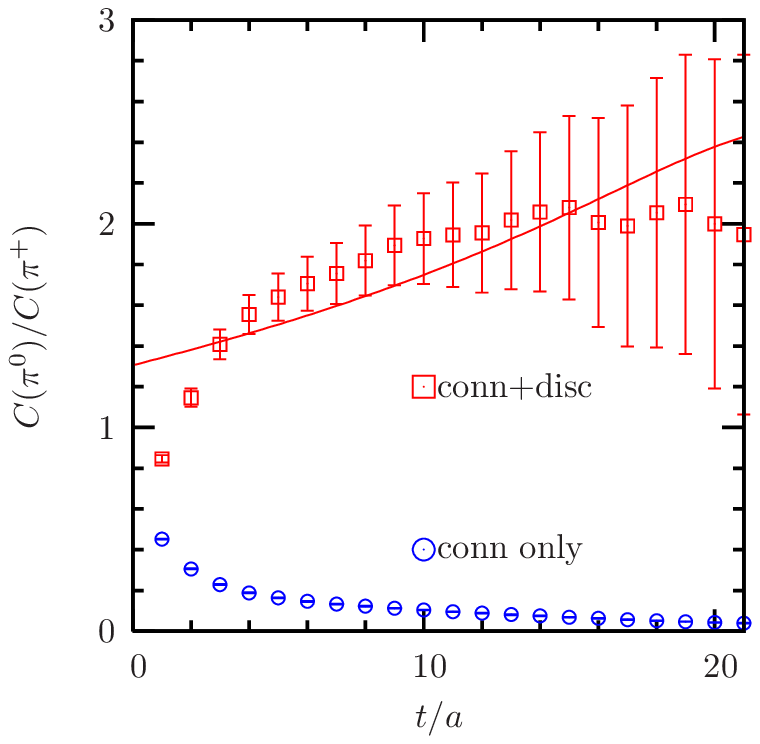} &
        \includegraphics[width=7cm]{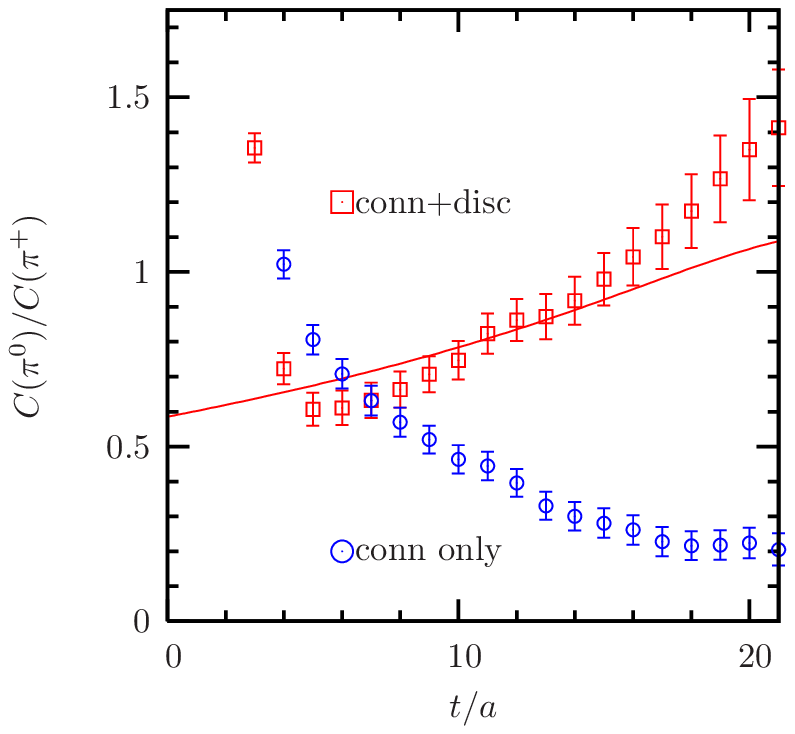}
   \end{tabular}
 \caption{Ratio of correlators between neutral and charged pions. The
left  plot is for local operators $\bar{\chi} I \chi$  for $\pi^0$ and 
$\bar{\chi} \gamma_5 \chi$ for $\pi^+$ and the right  plot for local
operators $ \bar{\chi} \gamma_4 \gamma_5 \tau_3 \chi$ for $\pi^0$ and 
$\bar{\chi} \gamma_4 \chi$ for $\pi^+$. The  curves give the ratio
arising from the mass difference determined by the full fit.
 }
 \label{fig:pionrat}
\end{center}
\end{figure}

\begin{figure}[ht]
\begin{center}
        \includegraphics[width=9cm]{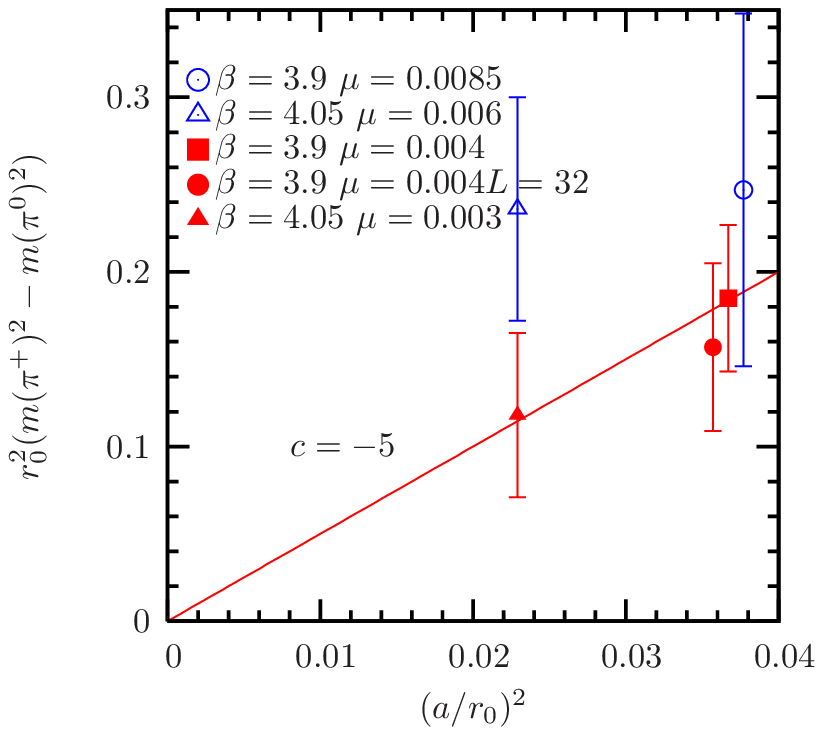} 
\end{center}
 \caption{Pion charge splitting}
 \label{fig:pion}
\end{figure}

For ETMC data with  $M(\pi^+)\approx 300$ MeV, $a=0.086$ fm and $La=2.1$ fm,
 we show the ratios of correlators in fig.~\ref{fig:pionrat}.
The disconnected pieces are seen to be relatively large - and reduce 
the charge splitting  as found previously~\cite{Jansen:2005cg}.  The $\pi^0$
is lighter than the $\pi^+$, unlike a  previous preliminary study of
dynamical fermions~\cite{Farchioni:2005hf}.
 From a $4 \times 4$ fit to these correlations, we obtain  $\Delta m
a=0.027(7)$.
  Since we expect $ r_0^2 ( m(\pi^0)^2 - m(\pi^+)^2) = c (a/r_0)^2 $,
 we compare this expression with  results from several 
lattice data sets~\cite{Urbach:2007aa} in fig.~\ref{fig:pion}.

   We see that, as expected, the  flavour splitting decreases as $a^2$. 
  The  sign and behaviour are consistent with Chiral PT and the nature
of the phase transition~\cite{Boucaud:2007uk} where $m(\pi^0)=0$. 
 We can use this determination to estimate the consequences of 
smaller lattice spacing, for instance less than 20\% pion splitting  for
$m(\pi^+)=200$ MeV provided $r_0/a> 8.2$.

\section {Flavour singlet PS meson: $\eta_2$}

\begin{figure}[hb]
\begin{center}
        \includegraphics[width=9cm]{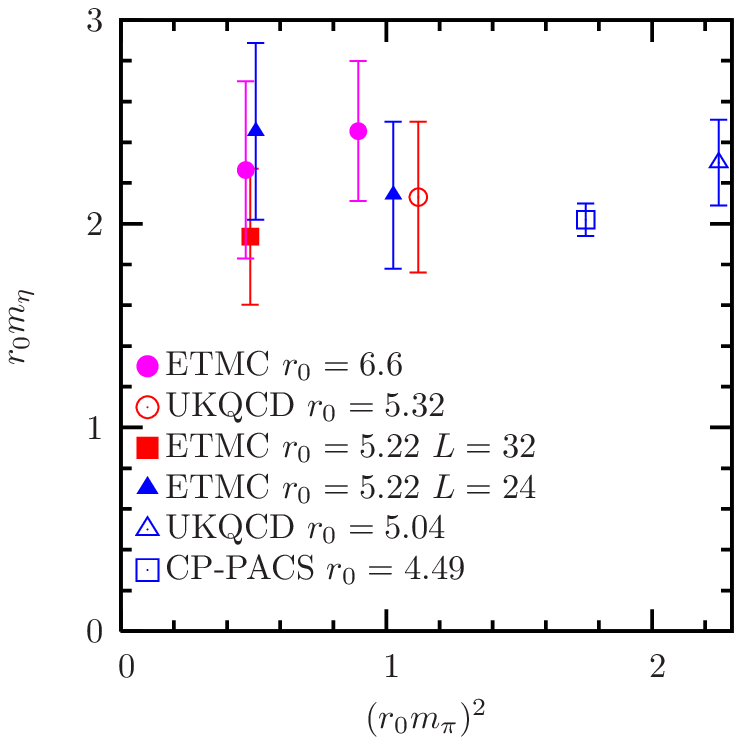} 
\end{center}
 \caption{$\eta_2$ mass versus quark mass.
Here we summarise results with light pions and $a < 0.1$ fm ($r_0/a> 4.5$).
 }
 \label{fig:eta}
\end{figure}

 In QCD, the flavour singlet pseudoscalar meson acquires a mass through
the  anomaly, so is not a goldstone boson. It is important to check that
this  feature, which is linked to topological charge fluctuations, is
reproduced in lattice evaluations. 
 With $N_f=2$ degenerate quarks, the flavour singlet pseudoscalar meson
(called $\eta_2$)  is related to the experimental $\eta'(958)$ and is
expected~\cite{McNeile:2000hf} to have a mass around 800 MeV. 
 We fit the $\eta_2$ correlators ($2 \times 2$ matrix with local and
non-local operator $\bar{\psi}\gamma_5 \psi  \to \bar{\chi}
\tau_3 \chi $ at maximal twist) for $t$-range 3-10 with 2 states.
 We compare  results from TMQCD~\cite{ETMClong} with older results (see
ref.~\cite{Allton:2004qq} for a review) in fig.~\ref{fig:eta}.
  Note that the ETMC results are at substantially smaller
quark masses.
 We see that the    $\eta_2$ mass is consistent with a constant
behaviour in the  chiral limit with $m(\eta_2) \approx .88$ GeV ($r_0
m(\eta_2)=2$).


 We now discuss why the errors are so large for the $\eta_2$, despite 
the fact that we
 measure all $t$ and $t'$, we use many gauge configurations   and stochastic
errors are small. The origin of the  problem is that the signal for  the
disconnected part of the correlator  comes from only a small part of the
total data sample.
 For instance (at $\mu=0.004$ with 48 $t$-values for 888 gauge
configurations) with $|t'-t|=10$,  2.1\% of the data contributes  26\%
of the signal. Thus the statistical impact of the data set is smaller
than expected since parts of the data  have big fluctuations (in a
fermionic loop related to topological charge density).
 So even more configurations are needed to get reliable and small errors
 in the case of disconnected  contributions.

\section{Vector mesons}


\begin{figure}
\begin{center}
   \begin{tabular}{l@{\hspace{1cm}}l}
        \includegraphics[width=7cm]{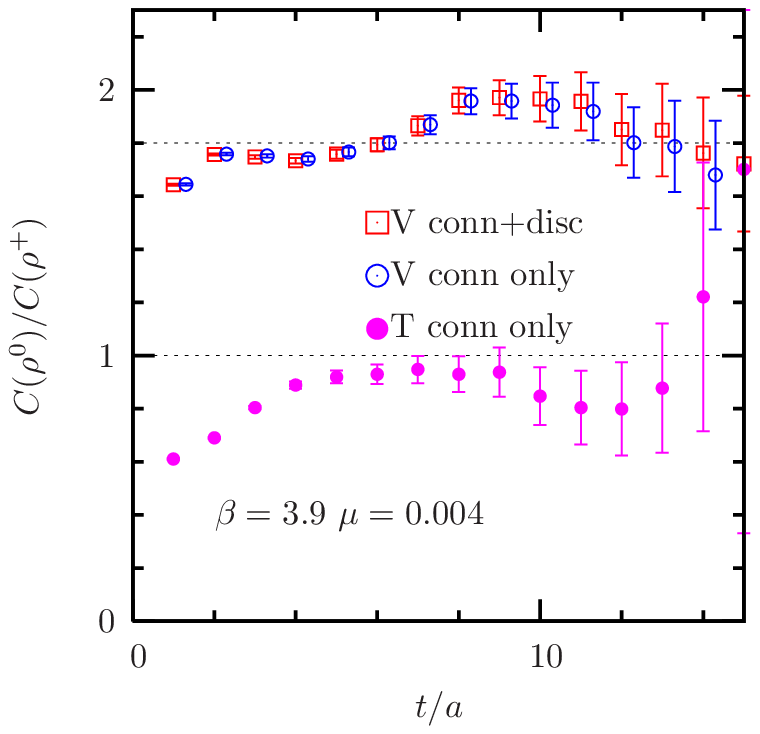} &
        \includegraphics[width=7cm]{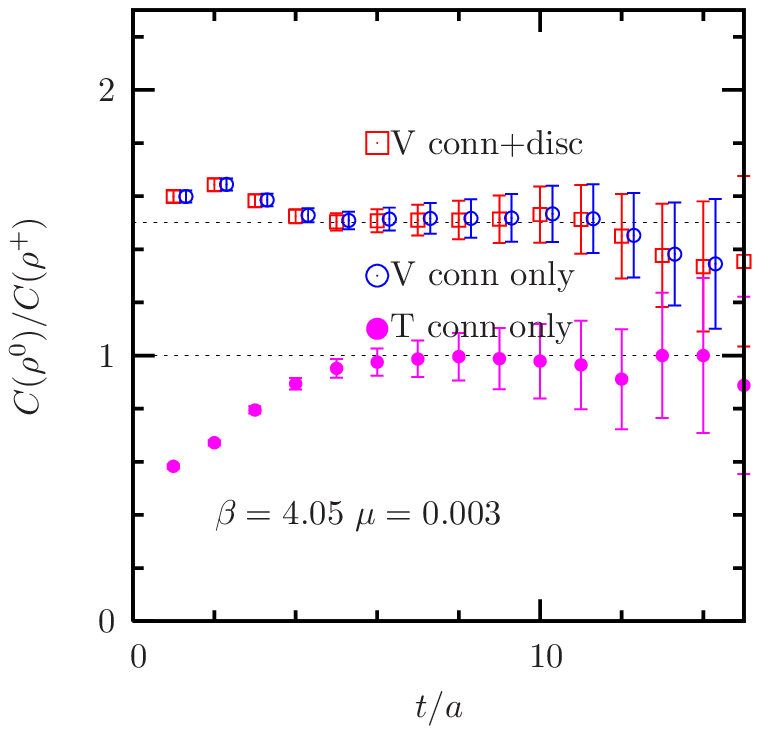}
   \end{tabular}
 \caption{Ratios of correlator for neutral $\rho$-mesons to charged. The
dotted lines guide the eye  in the case that there is no mass
splitting.}
 \label{fig:rho}
\end{center}
\end{figure}

 We compare the local-local correlators (including disconnected parts
for the neutral  meson) for vector mesons in fig.~\ref{fig:rho}.
  Note that the disconnected contribution is negligible for the 
vector coupling to neutral $\rho$-mesons.
 We find that the  ratio is consistent with constant (ie no  mass
splitting) and the value of that constant can be related  to
renormalisation constants. For the vector coupling to neutral 
$\rho$-mesons this implies $(Z_A/Z_V)^2 \approx 1.5$ at the finer
lattice spacing,  consistent with \cite{Dimopoulos:2007aa}. We find
agreement with a ratio of 1.0 for the tensor couplings, as expected
since there  is only one tensor renormalisation  which then cancels.

 From fits to a $2 \times 2$ matrix of correlators  (from the connected
neutral contribution only here in $t$-range 8-18) we obtain $\rho^0$
 masses (in lattice units).  We also report  values for the $\rho^+$
masses~\cite{ETMClong} (from fits to a $4 \times 4$  matrix of correlators). 
These values are consistent with no flavour splitting for vector mesons
as expected~\cite{Rossi:2007aa}.

 \begin{tabular}{llllll}
 $\beta$ & 3.9 & 3.9 & 3.9 & 4.05 & 4.05\\
  $\mu$ & $ 0.004$   &$ 0.004$   &  $ 0.0085$ & $ 0.003$ & $ 0.006$ \\
  $L$    & 32 & 24 & 24 & 32 &32 \\
 \hline 
 $am(\rho^0)$   & .400(25) & .395(17) & .419(17) & .372(29) & .346(12) \\
 $am(\rho^+)$   & .416(12) & .404(22) & .428(8) & .337(20) & .337(12)\\

\end{tabular}

\vspace{0.2cm}

 We now consider decay transitions from the vector meson to two pions
following the methods used in ref.~\cite{McNeile:2002fh}. The 
transition $\rho^0(0) \to \pi^+(1) \pi^-(-1)$, where 
$q(\pi)=\pm2\pi/L$, even for the lightest pion ($M(\pi^+)\approx 300$
MeV, $L=24a$, $a=0.086$ fm)  is not open (note, however, that decay  for
a $\rho^0$ meson with non-zero momentum is open).  For $\rho^0(0)$,
there is an energy splitting of $\Delta ma = 0.19$ assuming that the two
pion state  has twice the energy of a $\pi^+$ with appropriate momentum.


\begin{figure}[htb]
\begin{center}
        \includegraphics[width=9cm]{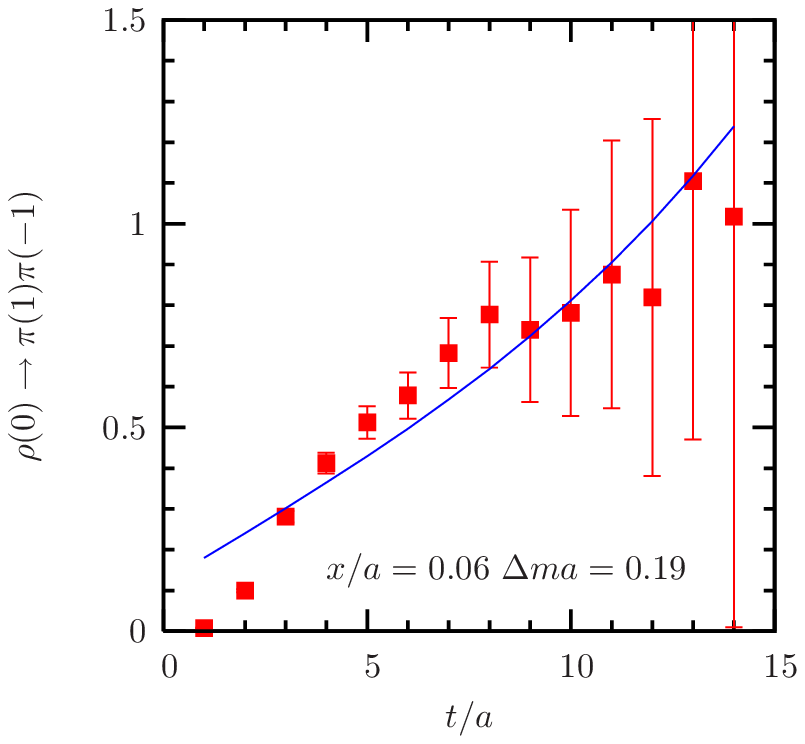}
\end{center}
 \caption{Transition $\rho^0 \to \pi^+ \pi^-$ from the lattice (with 
$M(\pi^+)\approx 300$ MeV, $a=0.086$ fm, $La=2.1$ fm ). The solid line 
is from a two state model ($\rho$ at rest and two pions with momentum
$q=\pm 2 \pi/L$) with energy gap $\Delta m a =2aE(\pi)-am(\rho)=0.19$
and transition amplitude $x/a$.
 }
 \label{fig:rhodecay}
\end{figure}

We show the normalised result in fig.~\ref{fig:rhodecay} where it is
compared to a two-state model~\cite{McNeile:2002fh}. The value of the
transition amplitude $x$ is consistent with the empirical $\rho$ decay
width.
 One can also estimate the effect of this mixing transition on the
$\rho$ mass  using the two-state model. This gives 
a downward shift of $ma$ of .02 (eg from .41 to .39 for the $\rho$
mass).
 This shift, induced by the proximity of the lightest two pion level, is
comparable to our statistical error in determining the  $\rho$ mass.  
 This suggests that we do not yet see major modifications of our $\rho$ 
meson masses from mixing with the decay channel.


\section {Flavour singlet scalar mesons}

 There is considerable confusion in allocating the experimental flavour-singlet
 scalar meson ($f_0$) spectrum to specific content: since
  scalar glueball,  $\bar{u}u+\bar{d}d$,  $\bar{s}s$,  $\pi \pi$ and/or
$KK$ in an S-wave, etc.,  can all contribute. Lattice QCD can help
considerably here, but it will be difficult as we  now illustrate.

 We took a first look in TMQCD   
 (here   $M(\pi^+)\approx 300$ MeV, $a=0.086$ fm, $La=2.1$ fm) 
and made a 2 state fit ($t$-range 6 to 23) to the 
 $ 6 \times 6 $ correlator matrix ($P,\ S,\ A_4$ both local and fuzzed
at  sink and source, including disconnected contributions).
 We find states
 $ma=.103(5)$ { ($\pi^0$)}  and $m'a=.227(28)$
{($f_0$, energy consistent with  $2 m(\pi^0)$)}.

 Thus in the scalar channel we find  a clear signal - but at the mass of two 
pions. 
 This is not unexpected - but emphasises the problems of studying 
scalar mesons with light quarks in dynamical lattice gauge theory, where the 
light two-body state will dominate the correlators.

\section{Summary}

   TMQCD allows efficient evaluation of disconnected contributions using
a powerful variance reduction method.

 The  $\pi$ charge splitting goes to zero like $a^2$ as expected, and
the sign is consistent with the nature of the phase transition.

The flavour singlet pseudoscalar meson ($\eta_2$) has been studied to
lighter quarks  than previously and is consistent with  a  mass of
around 800 MeV in chiral limit.

For vector mesons we find negligible charge splitting and 
the decay (transition to $\pi$ $\pi$) is  accessible to study.

 Our study of the flavour singlet scalar meson ($f_0$) with light
dynamical quarks,  finds  the expected  $\pi$ $\pi$ contribution which
will obscure further study of heavier states.


\providecommand{\href}[2]{#2}\begingroup\raggedright\endgroup

\end{document}